\documentclass[twocolumn]{cinc}
\usepackage[utf8]{inputenc}
\usepackage{graphicx}
\usepackage{amsmath} 
\usepackage{wrapfig}
\usepackage{multirow}
\usepackage{float}
\usepackage{subfig}
\usepackage{svg}
\usepackage{caption}
\usepackage{xurl}
\usepackage{algorithm}
\usepackage{program}
\usepackage{booktabs}

\usepackage{enumitem}
\setlist[enumerate,1]{label= (\arabic*),leftmargin=*,align=right}

\begin{document}
\title{Case Study: Fetal Breathing Movements as a Proxy for\\Fetal Lung Maturity Estimation}

\author{Márton Á. Goda$^{1}$, Ron Beloosesky$^{2}$, Chen Ben David$^{2}$, Zeev Weiner$^{2}$, Joachim A. Behar$^{1}$\\
 \ \\
$^1$Faculty of Biomedical Engineering, Technion, Israel Institute of Technology, Haifa, Israel
 \ \\
$^2$Rambam Health Care Campus, Haifa, Israel}

\maketitle

\begin{abstract}
Premature births can lead to complications, with fetal lung immaturity being a primary concern. Currently, fetal lung maturity (FLM) requires an invasive surfactant extraction procedure between the 32nd and 39th weeks of pregnancy. Unfortunately, there is no non-invasive method for FLM assessment. This work hypothesized that fetal breathing movement (FBM) and surfactant levels are inversely coupled and that FBM can serve as a proxy for FLM estimation. To investigate the correlation between FBM and FLM, antenatal corticosteroid (ACS) was administered to increase fetal pulmonary surfactant levels in a high-risk 35th-week pregnant woman showing intrauterine growth restriction. Synchronous sonographic and phonographic measurements were continuously recorded for 25 minutes before and after the ASC treatments. Before the ACS injection, 268 continuous movements FBM episodes were recorded. The number of continuous FBM episodes significantly decreased to 3, 43, and 79 within 24, 48, and 72 hours, respectively, of the first injection of ACS, suggesting an inversely coupled connection between FBM and surfactant level s. Therefore, FBM may serve as a proxy for FLM estimation. Quantitative confirmation of these findings would suggest that FBM measurements could be used as a non-invasive and widely accessible FLM-assessment tool for high-risk pregnancies and routine examinations.
\end{abstract}


\section{Introduction}

The World Health Organization (WHO) estimates that 15 million preterm deliveries occur worldwide each year, representing 5-18\% of neonates born globally \cite{Blencowe_2013}. One of the primary complications of prematurity is fetal lung immaturity, which can lead to fetal complications and death. Accurate assessment of fetal Fetal lung maturity (FLM) is therefore essential. During gestation, pulmonary surfactant ables neonatal breathing by increasing total lung capacity and lung compliance \cite{Kuroki_1998}. Antenatal corticosteroid (ACS) treatment enhances the expression of surfactant-associated proteins and accelerates the overall structural maturation of the lungs \cite{Jobe_2021}. ACS therapy is a standard to accelerate fetal lung maturation \cite{Committee_2017}. 
An optimal ACS administration-to-birth interval likely exists, however variations in study design limit identification of this interval from available evidence \cite{Mcdougall_2023}. Current methods of FLM assessment based on surfactant extraction are invasive and reliable only after 34 weeks of gestation \cite{Jobe_2021}, which may be too late for clinical purposes.

Quantitative ultrasound analysis is a modern non-invasive assessment technique that has been used to evaluate FLM\cite{Ahmed_2021}. However, these examinations are not routinely used because long-term ultrasound measurement for clinical practice is not well defined\cite{Keenan_2022}. Therefore, there is a need for a new non-invasive method that can accurately evaluate FLM. Fetal breathing movements (FBMs) are rhythmic contractions and relaxations of the diaphragm, with one contraction and relaxation of the diaphragm  defined as an episode \cite{kovacs_2023method}. A series of concatenated episodes form an epoch \cite{Goda_2020novel}. The intensity of FBMs becomes stronger and more frequent as the fetus matures\cite{Goda_2020novel}. Long-term assessment of fetal activity and FBMs can be performed using fetal phonocardiography (fPCG) and fetal electrocardiography (fECG) \cite{behar_2016}.

Previous studies have suggested that the number of FBMs is reduced following ACS therapy, indicating that increased FLM reduces FBM \cite{Mulder_2009}. Yet, in 2022 it has been also shown by Kumar \textit{et al.} showed that FBM increases the surfactant production in an \textit{in vitro} microfluidic alveolus \cite{Kumar_2022}, suggesting that FBM may also affect FLM through increased surfactant production. The relationship between FBMs and FLM in utero is still unclear.

The number of FBM episodes \cite{Pillai_1990} and surfactant production continuously increases \cite{Mendelson_1991} with the progression of pregnancy. We hypothesized that FBMs can serve as a useful estimate of FLM. We further hypothesized that FBM and surfactant levels have an inversely coupled relationship.


\begin{table*}[hbt!]
\fontsize{9pt}{12pt}
\selectfont
\begin{center}
\caption{Antenatal corticosteroids (ACS) injections during high-risk pregnancy.}\label{tab_01}
\begin{tabular}{ccccccccc}
& \textit{\textbf{Type}} & \textit{\textbf{Dose}} & \textit{\textbf{Freq.}} & \textit{\textbf{Route of admin.}} & \textit{\textbf{AFI}} & \textit{\textbf{mean MHR}} & \textit{\textbf{mean FHR}} & \textit{\textbf{Exam. time}} \\ \hline
\textbf{1st exam.} & Before ACS             & -       & -  & -     & AEDF  & 95 bpm  & 150 bpm & 20 minute\\ \hline
\textbf{2nd exam.} & 24h after 1$^{st}$ ACS & 12 mg   & x1 & I.M.  & AEDF+ & 90 bpm  & 160 bpm & 25 minute\\ \hline
\textbf{3rd exam.} & 48h after 1$^{st}$ ACS & 12 mg   & x2 & I.M.  & NEDF  & 90 bpm  & 169 bpm & 22 minute\\ \hline
\textbf{4th exam.} & 72d after 1$^{st}$ ACS & -       & -  & -     & NEDF  & 95 bpm  & 150 bpm & 24 minute\\ \hline
\end{tabular}
\end{center}
Intramuscular injection (I.M.) is a direct injection into a muscle. The amniotic fluid index (AFI) showed Absent End-Diastolic Flow (AEDF). After administration, the AFI improved and showed normal end-diastolic flow (NEDF).
\end{table*}

\section{Materials and Methods}

This study investigated data collected from a high-risk pregnant woman with a male fetus at 35 weeks of gestation. The woman was treated with ACS due to based intrauterine growth restriction (IUGR) as the fetal weight was estimated at 1800 g by ultrasound measurement, which is below the expected range for 35th gestational age. There was no indication of hypertension, thyroid problems, chronic kidney or lung disease in the pregnant woman, and she did not smoke during or prior to the pregnancy. The woman had breast cancer and received chemotherapy during pregnancy. 

Four measurements were conducted with medical support. The ethics approval for using retrospective de-identified data was granted by the Rambam Health Care Campus Institutional Ethics Committee under IRB number 0537-22-RMB. The first examination was performed immediately before the ACS injection. Follow-up measurements were  carried out 24, 48, and 72 hours after the first ACS injection.

Because of IUGR, the woman received two intramuscular (I.M.) 12-mg betamethasone injections, at a 24-hour interval. Before the ACS injection, the amniotic fluid index (AFI) showed absent end-diastolic flow (AEDF), which is an abnormal finding indicating poor fetal perfusion. After treatment, the AFI improved and showed Normal end-diastolic flow (NEDF), indicating improved fetal blood flow and oxygenation. Throughout the pregnancy, the woman did not receive any medication, such as anti-arrhythmic drugs, anticoagulants, or beta-blockers apart from the ACS injection. The ACS treatment was successful.

\begin{table}[hbt!]
\fontsize{9pt}{12pt}
\selectfont
\caption{Demographic of the pregnant.}\label{tab_02}
\begin{center}
\addtolength{\tabcolsep}{0pt} 
\begin{tabular}{ccccc}
\hline
\textbf{\begin{tabular}[c]{@{}c@{}}Maternal \\ Age\end{tabular}} & \textbf{\begin{tabular}[c]{@{}c@{}}Fetal\\ Age\end{tabular}} & \textbf{\begin{tabular}[c]{@{}c@{}}Fetal \\ Weight\end{tabular}} & \textbf{\begin{tabular}[c]{@{}c@{}}Fetal Head \\ Circum.\end{tabular}} &  \textbf{\begin{tabular}[c]{@{}c@{}}Fetus\\ Sex\end{tabular}} \\ \hline
26 y& 35 w & 1800 g & 31.2 cm & male\\ \hline
\end{tabular}
\end{center}
\end{table}

All measurements were taken between 12 PM and 1 PM, 2-3 hours after the woman's last meal of the subject and for a 26-year-old woman under normal conditions (see Table \ref{tab_02}). To investigate the FBMs, synchronous measurements were made using a Fetaphon2000 fetal phonocardiogram (fPCG) device and a Samsung 3D ultrasound device. The fPCG which is a passive, low-cost, portable heart rate-measuring device that can be used for long-term fetal monitoring, offers an opportunity to estimate fetal FBM at the single FBM episode level\cite{kovacs_2023method}. During the 25-minute examination, the doctors manually annotated the FBM epochs, and subsequently, the FBM episodes were detected during the post-processing stage. 

\begin{figure}[!hbt]
  \centering
  \includegraphics[width=\columnwidth]{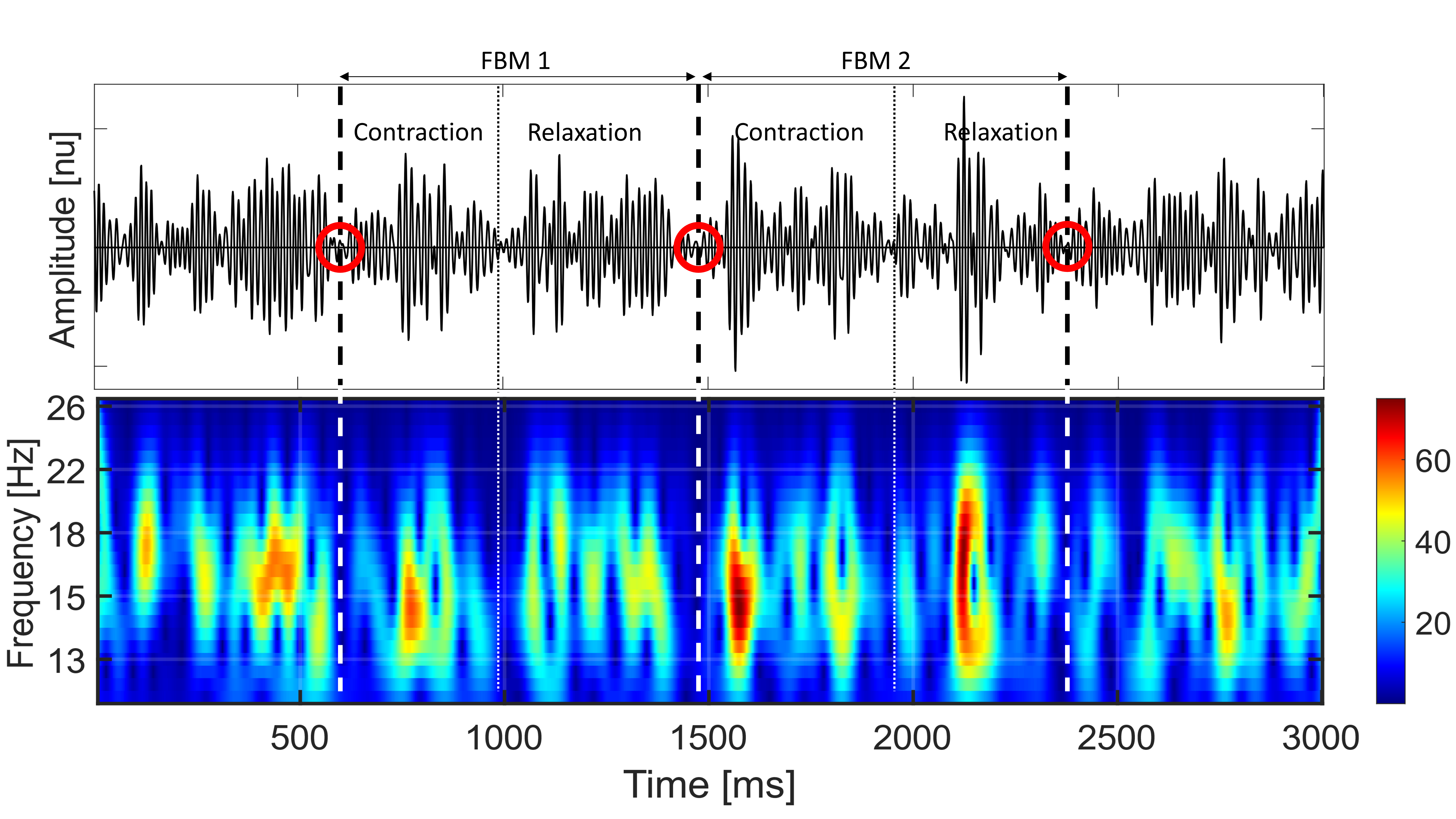}
  \caption{\scriptsize{This figure illustrates the FBM episodes in the fPCG signal. The signal shows a visible minimum point at the beginning of diaphragm contraction and relaxation. The dashed lines depict the starting point of diaphragm contraction and relaxation. The red circles indicate the beginning of the FBM episodes.}}
  \label{fig:FBM_sample}
\end{figure}

To assess the duration of each FBM episode individually, the time between the starting point of two consecutive episodes (dSP) was calculated. FBM episodes are defined as the time between two FBM start points (see Figure \ref{fig:FBM_sample}), and include the minimum zones \cite{Goda_2021phonography} since these points are detectable \cite{kovacs_2023method} using the ultrasound and the fPCG measurements. However, it is worth noting that there can be a gap between two episodes \cite{Goda_2020novel}. In the case of continuous FBMs, the dSP was expected to be less than 3 seconds. The mean (AVG), standard deviation (STD), median (MED) and the interquartile range (IQR) of the duration of FBM episodes were computed.

\begin{figure*}[!hbt]
  \centering
  \includegraphics[width=0.93\textwidth]{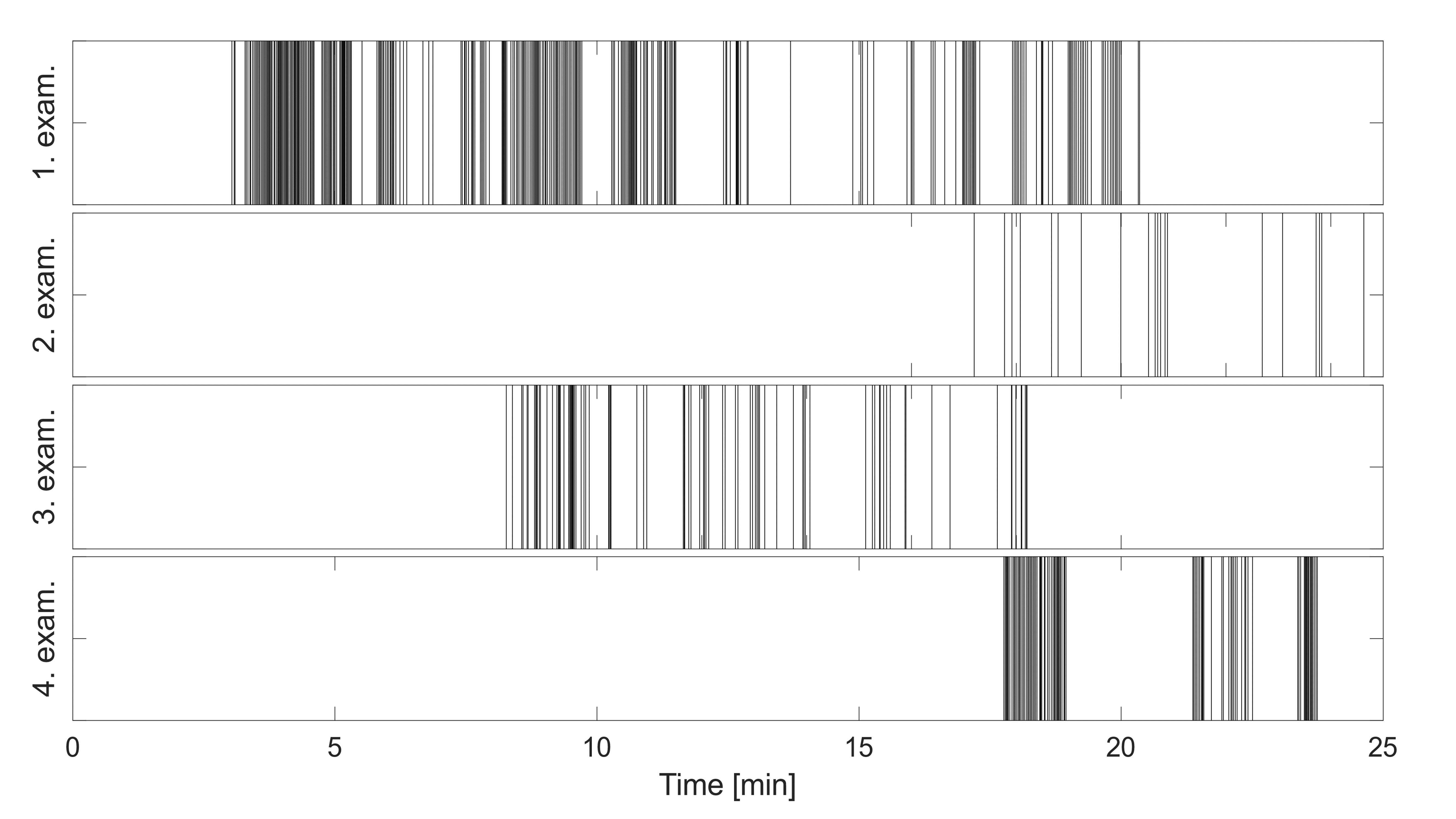}
  \caption{The frequency of fetal breathing movement (FBM) episodes during the different examinations. Each vertical line corresponds to the beginning of a single FBM.}
  \label{fig:FBMs}
\end{figure*}

\section{Results}
The frequency and number of FBM episodes varied throughout the four synchronous measurements. FBMs were the most frequent during the first examination, which was conducted before the ACS injection, and were observable almost throughout the entire examination. After the first ACS injection, the number of FBMs decreased significantly but increased again during the second and third examinations (see Figure \ref{fig:FBMs}).

\begin{figure}[!hbt]
  \centering
  \includegraphics[width=\columnwidth]{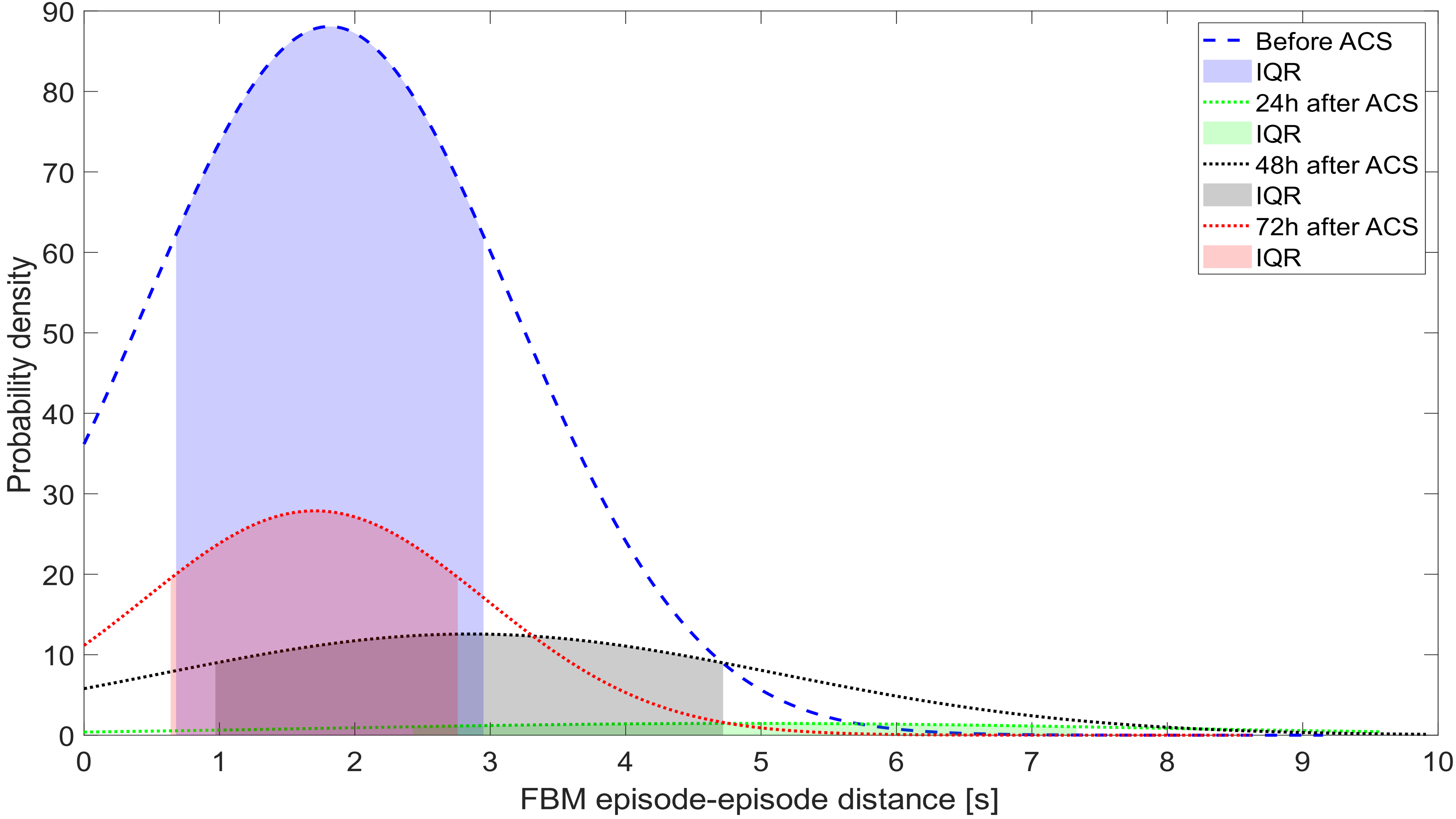}
  \caption{\scriptsize{This figure demonstrates the impact of the ACS injection on FBM episodes. The dashed lines in the graph depict the probability density of the time between two successive FBM episodes over the 25-minute examination periods, while the shaded area corresponds to the interquartile range (IQR) of the episode distances. Notably, following the first injection, the median FBM episode distance decreased, but after 72 hours, it returned to levels similar to those observed in the first examination.}}
  \label{fig:ACS-FBM}
\end{figure}

Immediately before the first ACS injection, 319 FBM episodes were recorded (see Table \ref{tab_03}). Assuming that the time between consecutive FBM episodes is less than 3 seconds  for continuous FBM activity, we can count 268 continuous FBM episodes, which accounts for 84\% of the total number of FBMs during the first examination. Following the first ACS injection, the number of continuous FBM episodes decreased drastically, with only three episodes (15\% of the FBM episodes of the second examination) observed 24 hours after dosing (see Figure \ref{fig:ACS-FBM}). However, 48 hours after first injection (24 hours after the second injection), the number of FBMs increased to 43 visible episodes of continuous FBM (49.4\% of the FBM episodes of the third examination). Finally, 72 hours after the first ACS injection, 79 episodes of continuous FBM were observed (87.7\% of FBM episodes of the fourth examination).

\begin{table}[hbt!]
\fontsize{9pt}{12pt}
\selectfont
\caption{FBM episodes and ACS injections.}\label{tab_03}
\begin{center}
\addtolength{\tabcolsep}{-2.7pt}    
\begin{tabular}{lcccc}
&\multicolumn{4}{c}{Number of FBM episodes}\\
& \textbf{Before ACS} & \textbf{After 24h} & \textbf{After 48h} & \textbf{After 72h} \\
& 318 & 20 & 87 & 90\\ \cline{2-5}
\textit{dSP \textless{}3s} & 268 & 3  & 43 & 79 \\ \cline{2-5}

\multicolumn{5}{c}{}\\
\textbf{} &\multicolumn{4}{c}{Length of FBM episodes (\textit{dSP \textless{}3s)}}\\
& \textbf{Before ACS} & \textbf{After 24h} & \textbf{After 48h} & \textbf{After 72h} \\
\textit{\textbf{AVG}}      & 1.42 s & 2.09 s & 1.29 s & 1.38 s\\ \cline{1-5}
\textit{\textbf{STD}}      & 0.49 s & 0.17 s & 0.67 s & 0.52 s\\ \cline{1-5}
\textit{\textbf{MED}}      & 1.33 s & 2.81 s & 1.22 s & 1.35 s\\ \cline{1-5}
\textit{\textbf{IQR}}      & 0.68 s & 0.17 s & 0.83 s & 0.71 s\\ \cline{1-5} 
\end{tabular}
\end{center}
\end{table}

The median distances between two consecutive FBM episodes were 1.8 seconds  and 1.7 seconds  for the first and fourth measurements, respectively. In contrast, for the second and third measurements, the median distances were 2.9 seconds  and 4.9 seconds, respectively,  indicating slower FBMs with higher gaps between episodes.

\section{Conclusion and Discussion}

In this case analysis, the number of the FBMs decreased during ACS treatment, but subsequently continuously increased. This suggests an inversely coupled connection between FBM and pulmonary surfactant level. Namely, if the surfactant level is too low, the number of FBM episodes may increase, and if the surfactant levels are too high, the number of FBM episodes may decrease until reaching an equilibrium state. Therefore, in the absence of an invasive surfactant measurement, FBM can serve as a reliable proxy for FLM estimation.

Non-invasive FLM assessment can have a high a considerable on clinical practice by providing for safe investigation of high-risk pregnancies. FLM assessment can support ACS therapy in the early stage of the last trimester when such treatment is clinically challenging in clinical practice \cite{Mulder_2009}. Furthermore, fPCG enables long-term remote fetal measurement \cite{Kahankova_2022}, which could offer an accurate telemedicine based assessment of FLM at any time during the third trimester.

\section{Acknowledgements}

MAG and JAB acknowledge the Faculty of Biomedical Engineering, Technion, Israel Institute of Technology. RB, CBD, and ZW  acknowledge the Rambam Health Care Campus.

\bibliographystyle{cinc}
\bibliography{new_refs}

\begin{correspondence}

Márton Á. Goda, PhD\\
marton.goda@campus.technion.ac.il\\
AIMLab, Technion-IIT, Haifa, 32000, Israel\\
\end{correspondence}
\end{document}